\documentclass[superscriptaddress,twocolumn,showpacs,amsmath,amssymb]{revtex4}

\usepackage{graphicx}
\usepackage{dcolumn}
\usepackage{bm}

\input epsf
\renewcommand{\vec}[1]{\mbox{\boldmath $#1$}}

\begin{document}

\preprint{}

\title{
Effect of electronic environment on 
neutrino-nucleus reactions at r-process sites 
}

\author{F. Minato}
\affiliation{
Department of Physics, Tohoku University,
Sendai 980-8578, Japan}

\author{K. Hagino}
\affiliation{
Department of Physics, Tohoku University,
Sendai 980-8578, Japan}

\author{N. Takigawa}
\affiliation{
Department of Physics, Tohoku University,
Sendai 980-8578, Japan}

\author{A. B. Balantekin}
\affiliation{Department of Physics, University of Wisconsin, 
Madison, Wisconsin 53706}

\author{Ph. Chomaz}
\affiliation{
GANIL-CEA/DSM-CNRS/IN2P3, 
B.P. 55027, 
F-14076 Caen C\'edex 5,
France}

\date{\today}

\begin{abstract}
We discuss effects of the electron plasma on 
charged-current neutrino-nucleus reaction, $(\nu_e,e^-)$
in a core-collapse supernova environment. 
We first discuss 
the electron screening effect on the final state 
interaction between the outgoing electron 
and the daughter nucleus. 
To this end, 
we solve the Dirac equation for the outgoing electron with 
the screened Coulomb potential obtained with 
the Thomas-Fermi approximation. 
In addition to the screening effect, we also 
discuss the Pauli blocking effect 
due to the environmental electrons on the spectrum of the 
outgoing electron.  
We find that both effects hinder the cross section of 
the charged-current reaction, especially at 
low incident energies. 
\end{abstract}

\maketitle

\section{Introduction} 
A large number of neutrinos are emitted from a core-collapse 
supernova. 
These neutrinos interact with nuclei through the weak interaction. 
Although their cross sections are small, 
it is agreed that their contribution to 
nucleosynthesis (that is, r-, $\nu$- and p-processes) 
is not negligible due to the large neutrino luminosity 
\cite{Balantekin:2003ip}. 
Neutrinos may even play a leading role in some cases. 
For instance, 
Yoshida {\it et al.} recently argued that the abundance ratio 
between $^7$Li and $^{11}$B is sensitive to the 
$\nu$-process and thus can be used to extract information on 
the neutrino mass hierarchy \cite{Yoshida2006}.
Also, the abundance ratio between U and Th elements, 
which has been used as 
a cosmochronometer, may be affected by the $\nu$-process.  
It is thus important 
to calculate with high accuracy 
the cross section of the neutrino-nucleus reactions 
in a dense star. 

In the supernova nucleosynthesis, only the charged current reactions
of the electron neutrinos, 
$\nu_{e}$, and the electron anti-neutrinos, $\overline{\nu}_e$, 
are relevant, since those of $\nu_{\mu}$ and $\nu_{\tau}$ (and their 
antineutrinos) are 
suppressed due to the threshold effects. 
These processes are represented by
\begin{eqnarray}
\nu_{e} + (Z,A) &\rightarrow& e^- + (Z+1,A),\label{eq:1} \\
\overline{\nu}_{e} + (Z,A) &\rightarrow& e^+ + (Z-1,A).\label{eq:1-2} 
\end{eqnarray}
The average energies of $\nu_{e}$ and $\overline{\nu}_{e}$ neutrinos 
in a core collapse supernova are typically 11 MeV and 16 MeV, 
respectively \cite{QFMMWW93}. 
It must be emphasized, however, that the 
exact hierarchy of neutrino energies depend on the details of the 
underlying microphysics \cite{micro}. 
The outgoing electron and positron feel the Coulomb interaction 
from the daughter nucleus as they leave. 
This final state interaction affects the neutrino-nucleus 
reaction rate \cite{Morita,Walecka}. 

In the supernova environment, the motion of the outgoing electron is 
further perturbed by 
other charged particles in the environment besides the daughter
nucleus, such as 
protons, electrons, $\alpha$ particles and other heavier nuclei.
Such effects have been considered in Ref. \cite{Itoh2002} for 
electron capture rates in a dense star.  
Furthermore, in a high electron density, 
the reaction (\ref{eq:1}) is suppressed 
because low energy electron states are Pauli blocked. 
It is crucial to take into account those two effects in order to 
accurately estimate the neutrino-nucleus reaction rate for
nucleosynthesis. 

In this paper, we perform such calculations, taking into 
consideration both the electron screening and Pauli blocking 
effects. 
For this purpose, we use the 
Thomas-Fermi theory for 
electron screening, 
which has often been used in many-body physics. 
We then calculate the cross section of a neutrino-nucleus reaction 
using the distorted wave Born approximation (DWBA) method, including 
the screening and the Pauli blocking effects simultaneously. 

The paper is organized as follows. In Sec. II, we detail our model for 
the charged current neutrino-nucleus reaction in a supernova
environment, that uses the Thomas-Fermi theory 
for electron
screening, and the DWBA method with the Pauli blocking effect. 
In Sec. III, we present the results of our calculation for the 
$^{56}$Fe($\nu_{e},e^-)^{56}$Co and $^{208}$Pb($\nu_{e},e^-)^{208}$Bi 
reactions, and discuss the effects of the electronic environment. 
We summarize our paper in Sec. IV. 

\section{Neutrino-nucleus reactions at finite electron density}
\subsection{Electron screening effect}

We first discuss the electron screening effect on the final state 
interaction between the outgoing electron and the daughter nucleus. 
Despite of the charge neutral condition of the environment, 
the screening effect can still be significant because the daughter nucleus 
attracts the environmental electrons and 
polarizes the background charge distribution. 
The positive charge particles are
relatively heavy, and thus we neglect its polarization.

We assume that the electron charge distribution is homogeneous with 
density $\rho_e^0$ in the absence of the daughter nucleus. 
This charge distribution of the environmental electrons 
is modified to $\rho_{e}(\vec{r})$ due to the presence of the 
daughter nucleus, whose charge distribution is represented by 
$e\rho_{\rm N}(\vec{r})$. 
Neglecting the 
the interaction among the environmental electrons, which is valid 
at high temperatures, the Coulomb field $\phi(\vec{r})$ at $\vec{r}$
from the daughter nucleus reads 
\begin{equation}
\phi(\vec{r})=\int\,d\vec{r}'\frac{e\rho_{\rm N}(\vec{r}')
-e\,\delta\rho_e(\vec{r}')}
{|\vec{r}-\vec{r}'|}, 
\label{eq:3}
\end{equation}
where $\delta\rho_e(\vec{r})\equiv \rho_e(\vec{r})-\rho_e^0$ is the
polarization charge.
In order to evaluate this function, we assume a sharp-cut charge 
distribution for $\rho_{\text{N}}$,  that is, 
$\rho_{\text{N}}(\vec{r})=[3Z/(4\pi R^3)]\cdot\theta(R-r)$ 
with a nuclear radius of $R$. 
Here, $Z$ is the atomic number of the 
daughter nucleus.  

For the electron density $\rho_e$,
we use the Thomas-Fermi theory. 
The polarization charge then reads 
\begin{equation}
\delta\rho_{e}(\vec{r})=\frac{1}{3\pi^2\hbar^3}(2m\epsilon_{F}(\vec{r}))^{3/2}
-\rho_e^0. 
\label{TF}
\end{equation}
Here, $m$ is the electron mass, and the local Fermi energy 
$\epsilon_{F}(\vec{r})$ is given by 
\begin{equation}
\epsilon_{F}(\vec{r})=\epsilon_{F}^{0}+e\phi(\vec{r}), 
\end{equation}
with
$\epsilon_{F}^{0}=(3\pi^2\hbar^3\rho_{e}^{0})^{2/3}/2m$. 
We have assumed that atoms are 
completely ionized, without any bound electrons. 

Since the electron density 
$\rho_e$ depends explicitly on the Coulomb field $\phi$, 
we evaluate 
Eq. (\ref{eq:3}) self-consistently, 
assuming the spherical symmetry.
The boundary condition is imposed so that
the Coulomb potential vanishes at the radius 
where the net negative charge inside is 
equal to the charge number of the daughter nucleus. 

Once the Coulomb field $\phi$ is obtained, we solve the Dirac 
equation for the outgoing election with the potential
$V_C(r)=-e\phi(r)$. 
Writing the wave function of the outgoing electron as, 
\begin{equation}
\psi^E_{jlm}(\vec{r})=\frac{1}{r}
\left(
\begin{array}{c}
iP^E_{jl}(r)\,{\cal Y}_{jlm}(\hat{\vec{r}}) \\
Q^E_{jl}(r)\,{\cal Y}_{j,2j-l,m}(\hat{\vec{r}}) 
\end{array}
\right),
\end{equation}
with 
\begin{equation}
{\cal Y}_{jlm}(\hat{\vec{r}}) 
=\sum_{m_l,m_s}\langle l\,m_l\,1/2\,m_s|jm\rangle 
Y_{lm_l}(\hat{\vec{r}})\,\chi_{m_s},
\end{equation}
where $\chi_{m_s}$ is the two component spin wave function, 
the Dirac equation reads 
\begin{eqnarray}
\frac{d P^E_{jl}}{dr}&=&-\frac{\kappa}{r}\,P^E_{jl}(r)
-[E+m-V_c(r)]\,Q^E_{jl}(r), \\
\frac{d Q^E_{jl}}{dr}&=&\frac{\kappa}{r}\,Q^E_{jl}(r)
+[E+m-V_c(r)]\,P^E_{jl}(r), 
\end{eqnarray}
where $\kappa=(l-j)(2j+1)$. 
We use the computer code RADIAL \cite{SFW95} to solve the 
Dirac equation. 
We use these wave functions in Sec. II C 
in order to compute the cross sections of charged current
neutrino-nucleus reactions. 

\subsection{Pauli blocking effect}

In the high density limit, 
the environmental electrons are almost degenerate and 
the occupation probability of the lowest energy states 
is close to unity.  
In this situation, the 
production of a low energy electron is suppressed 
due to the Pauli blocking effect of the environmental electrons. 
For a given electron density $\rho_{e}^{0}$, the Fermi energy $\mu_e$ 
can be estimated using the relativistic Fermi gas model,
\begin{equation}
\rho_{e}^{0}=\frac{m^3c^6}{\pi^2\hbar^3}\int^{\infty}_{0}
\frac{\sinh^2x
  \cosh x}{\exp(\beta(mc^2\cosh x-\mu_{e}))+1}
\,dx,
\label{eq:6}
\end{equation}
where $\cosh x=p/mc^2$, $p$ being the momentum of the electron.
In order to take into account the Pauli blocking effect, we multiply 
a factor $(1-f_{e}(E_{e},T_{e},\mu_{e}))$ to the cross section, where 
$E_e$ is the energy of the outgoing electron and 
$f_{e}$ is the distribution function of 
the environmental electrons given by 
\begin{equation}
f_{e}(E,T_{e},\mu_{e})
=\frac{1}{1+\exp[(E-\mu_e)/T_e]}. 
\end{equation}

\subsection{Cross sections}

We estimate the cross sections of the neutrino-nucleus reactions using 
the DWBA method, which is the most natural way to include the 
screening correction. This method in fact improves 
methods typically used to calculate final state interactions, such as 
so-called 
Fermi function and effective momentum approximations
\cite{Engel1998}. 
The cross sections for the Fermi type transition from the ground state 
of a parent nucleus with $J_i^\pi=0^+$ to the 
$J^\pi$ state in the daughter nucleus is given by 
\cite{Engel1998}, 
\begin{equation}
\begin{split}
&\frac{d\sigma}{dE_{e}}= 
G^2\frac{E_{e}+k_{e}}{2k_{e}}\Big(1-f_{e}(E_{e},T_{e},\mu_{e})\Big)\\
&\times\sum_{j,L}\left(
\begin{split}
&(2j+1)(2L+1)\\
&\times\left(
\begin{split}
&\langle j-\frac{1}{2}0,L0|J0\rangle^2(F_{LJ,j}^{-2}+G_{LJ,j}^{+2})\\
&+\langle j+\frac{1}{2}0,L0|J0\rangle^2(F_{LJ,j}^{+2}+G_{LJ,j}^{-2})
\end{split}
\right)\\
&-2\langle j+\frac{1}{2}0,L0|J0\rangle\\
&\times\left(
\begin{split}
&\sqrt{[J^2-(L+\frac{1}{2}-j)^2]}\\
&\times\sqrt{[(J+1)^2-(L+\frac{1}{2}-j)^2]}\\
&\times\langle j-\frac{1}{2}0,L+10|J0\rangle\\
&\times(F_{L+1J,j}^{-}G_{LJ,j}^{-}-F_{LJ,j}^{+}G_{L+1J,j}^{+})\\
&+\sqrt{[J^2-(L+\frac{1}{2}+j)^2]}\\
&\times\sqrt{[(J+1)^2-(L+\frac{1}{2}+j)^2]}\\
&\times\langle j-\frac{1}{2}0,L-10|J0\rangle\\
&\times(F_{L-1J,j}^{-}G_{LJ,j}^{-}-F_{LJ,j}^{+}G_{L-1J,j}^{+})\\
\end{split}
\right)
\end{split}
\right), 
\end{split}
\label{DWBA}
\end{equation}
where $G$ is the Fermi constant. 
$E_{e}$ and $k_{e}$ are the energy and momentum of the 
outgoing electron, respectively, and 
$k_{\nu}$ is the incident neutrino momentum.
The relation between the neutrino energy $E_{\nu}$ and the 
electron energy $E_{e}$ is 
$E_{\nu}=\hbar k_\nu c=E_{\text{x}}+E_{e}-Q$, where $E_{\rm x}$ is the 
excitation energy for the nuclear transition with respect to the 
ground state of the parent nucleus, and $Q=m_{n}c^2-m_{p}c^2$ is the 
mass difference between neutron and proton. 
$F_{LJ,j}^{\pm}$ and $G_{LJ,j}^{\pm}$ are defined as \cite{Engel1998}, 
\begin{eqnarray}
F_{LJ,j}^{\pm}&=&\int d\vec{r}\,\frac{P_{j,j\pm1/2}^{E_{e}}(r)}{r}\,
j_{L}(k_{\nu}r)Y_{JM}(\hat{\vec{r}})\,\rho_{fi}(\vec{r})\\
G_{LJ,j}^{\pm}&=&\int d\vec{r}\,\frac{Q_{j,j\pm1/2}^{E_{e}}(r)}{r}\,
j_{L}(k_{\nu}r)Y_{JM}(\hat{\vec{r}})\,\rho_{fi}(\vec{r}),
\end{eqnarray}
respectively, 
$\rho_{fi}(\vec{r})$ being the transition density for the nuclear Fermi
transition. 

\begin{figure}[h!]
\begin{center}
\includegraphics[width=0.91\linewidth, clip]{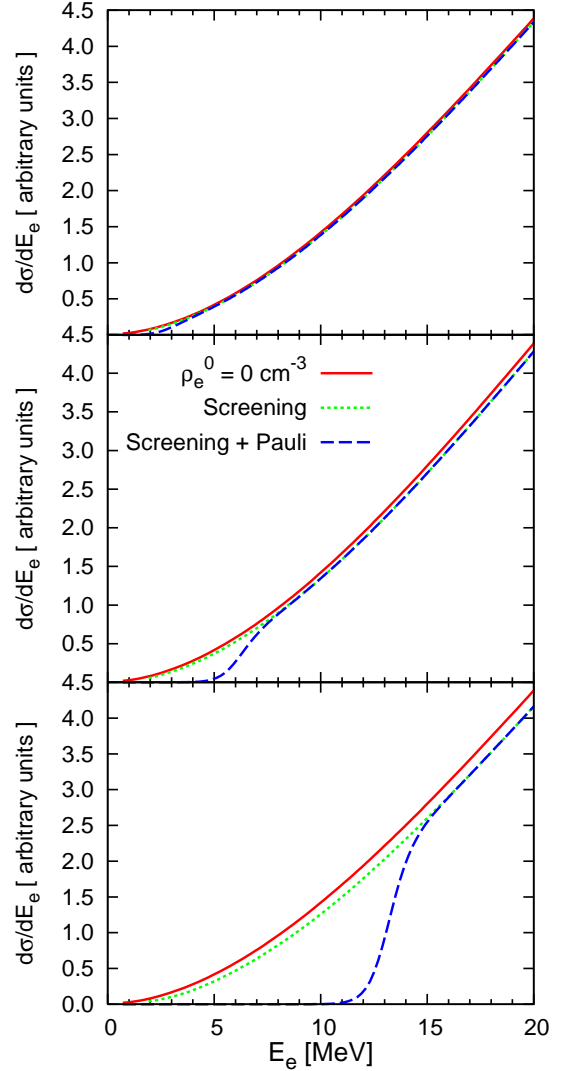}
\caption{
(Color online) 
The cross sections for the charged current 
$\nu_e + ^{56}$Fe $\rightarrow e^- + ^{56}$Co reaction 
as a function of the energy of the outgoing electron 
for three densities of the environmental electrons, 
$\rho_{e}^{0}$. 
The Fermi transition to the $J^{\pi}=0^{+}$ state at $E_x=3.5$ MeV 
in $^{56}$Co is considered. 
The top, middle, and bottom panels are for 
$\rho_{e}^{0}= 10^{32}, 10^{33}$, and 
10$^{34}$ cm$^{-3}$, respectively. 
The solid line shows the results in the absence of the environmental 
electrons, while the dotted line includes the electron 
screening effects. The dashed line takes into account both the 
screening and the Pauli blocking effects. }
\label{fig:1}
\end{center}
\end{figure}
\begin{figure}[h!]
\begin{center}
\includegraphics[width=0.91\linewidth, clip]{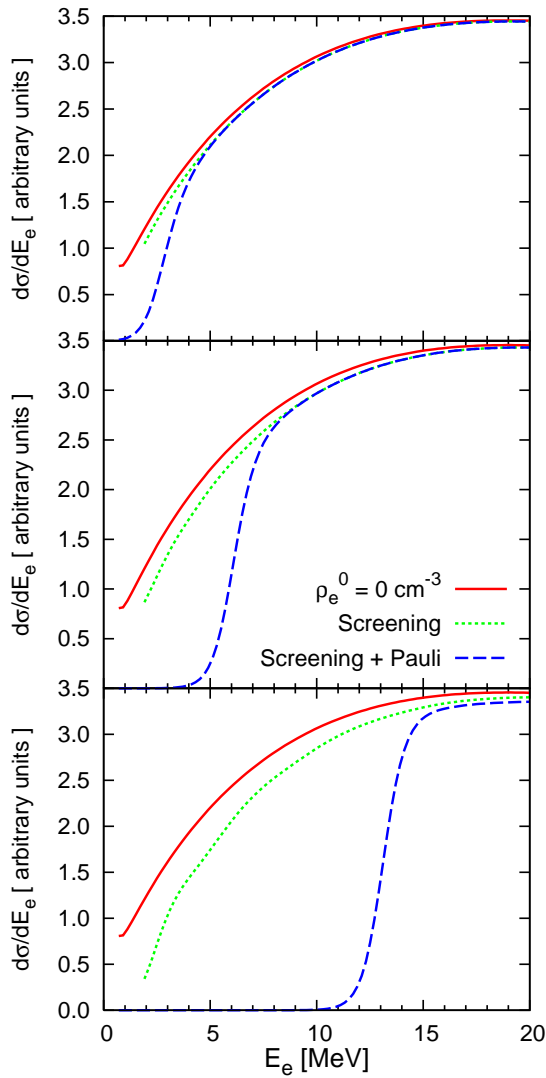}
\caption{
(Color online) 
The same as Fig. 1, but for the 
$\nu_e + ^{208}$Pb $\rightarrow e^- + ^{208}$Bi reaction.
The Fermi transition to the $J^{\pi}=0^{+}$ state at $E_x=15.0$ MeV 
in $^{208}$Bi is considered. }
\label{fig:2}
\end{center}
\end{figure}

\section{Results}

We now evaluate numerically the electron screening and the Pauli 
blocking effects on the charged current 
$^{56}$Fe$(\nu_{e},e^{-})^{56}$Co 
and $^{208}$Pb$(\nu_{e},e^{-})^{208}$Bi 
reactions. 
We set the electron temperature to be $T_{e}$ = 0.5 MeV. 
We consider the Fermi type transition to the
$J^{\pi}=0^{+}$ state at
$E_{\text{x}}=$ 3.5 MeV in $^{56}$Co \cite{Murakami1983} and
$E_{\text{x}}=$ 15.0 MeV in $^{208}$Bi \cite{Krasznahorkay2001}.
For simplicity, we follow Ref. \cite{Engel1998} and 
assume the transition density which is proportional
to $\rho_{fi}\propto\delta(r-R)Y_{JM}(\theta,\phi)$.

The differential cross sections $d\sigma/dE_e$ for 
the $^{56}$Fe$(\nu_{e},e^{-})^{56}$Co 
and the $^{208}$Pb$(\nu_{e},e^{-})^{208}$Bi 
reactions are shown in Figs. 1
and 2, respectively. The solid line shows the results in the absence
of the environmental electrons. The top, middle and bottom panels are 
for the electron density of 
$\rho_{e}^{0}$ = 10$^{32}, 10^{33}$, and 10$^{34}$ cm$^{-3}$, 
respectively. 
The dotted line denotes the results with the electron 
screening effects, while the dashed line takes into account both the 
screening and the Pauli blocking effects. For the electron density 
smaller than $10^{31}$ cm$^{-3}$, we find that both the effects 
are marginal. The screening effect is larger in 
the $^{208}$Pb$(\nu_{e},e^{-})^{208}$Bi than in the 
$^{56}$Fe$(\nu_{e},e^{-})^{56}$Co reaction, as is expected. 
The electron Fermi energy obtained with Eq. (\ref{eq:6}), for 
$\rho_{e}^{0}$ = 10$^{32}, 10^{33}$, and 10$^{34}$ cm$^{-3}$ is 
2.59, 5.99, and 13.1 MeV, respectively. 
We see from the figures that the Pauli blocking effects are
important below these energies. 
We have confirmed that the screening and the blocking effects
disappear in the high $E_e$ limit. 

For the anti-neutrino induced reactions, (\ref{eq:1-2}), 
the screening effect increases the cross section in contrast to the 
neutrino induced reactions shown in Figs. 1 and 2, while the Pauli blocking 
effect is irrelevant in the absence of a 
positron background. The net effect of the environmental electrons 
thus increases the cross sections for these reactions. 

We next discuss total cross sections.  
In order to compute the total cross sections, we 
integrate the differential cross sections, Eq. (\ref{DWBA}), with a 
weight factor given by the energy distribution for the incident neutrino, 
$n_\nu(E_\nu)$. That is, 
\begin{equation}
\sigma=\int_{0}^{\infty}
\frac{d\sigma}{dE_{\nu}}n_{\nu}(E_{\nu})\,dE_{\nu}. 
\end{equation}
Following Ref. \cite{Qian1995}, 
we assume that the distribution function $n_\nu$ is 
given by, 
\begin{equation}
n_{\nu}(E_{\nu})=\frac{1}{FT_{\nu}^3}\,
\frac{E_{\nu}^2}{e^{E_{\nu}/T_{\nu}}+1}, 
\label{Edistribution}
\end{equation}
with
\begin{equation}
F=\int_{0}^{\infty} dE\,\frac{E^2}{e^{E}+1}, 
\end{equation}
at distances outside a neutrino sphere, where r-process is considered to 
take place. Here $T_\nu$ is the neutrino temperature at the 
neutrino sphere. 

Figures 3 and 4 show the total cross sections for the 
$^{56}$Fe$(\nu_{e},e^{-})^{56}$Co 
and the $^{208}$Pb$(\nu_{e},e^{-})^{208}$Bi reactions as a function of
the density of the environmental electrons, respectively. These are
plotted as the ratio to the total cross sections in the absence of the 
environmental electrons, $\sigma_0$. The neutrino temperature $T_\nu$
in Eq. (\ref{Edistribution}) is set to be 4 MeV. 
The dotted line takes into account only the screening effects, while
the dashed line includes both the screening and the Pauli blocking
effects. 
We see that the Pauli blocking effect influences 
the $^{208}$Pb$(\nu_{e},e^{-})^{208}$Bi reaction much more 
significantly than 
the $^{56}$Fe$(\nu_{e},e^{-})^{56}$Co reaction. This is due to 
the fact that the neutrino energy distribution $n_\nu$ has a peak 
around $E_\nu=$ 8.87 MeV at the neutrino temperature considered here. 
For the former reaction, the excitation 
energy $E_x$ is 15 MeV and the energy of the produced electron is 
always small, for which the blocking effect is important. 
For the latter reaction, on the other hand, the 
excitation energy is relatively small ($E_x$=3.5 MeV), and the 
average energy of the outgoing electron is above the electron 
Fermi energy unless the electron density $\rho_e^0$ is large. 

\begin{figure}
\begin{center}
\includegraphics[width=0.91\linewidth, clip]{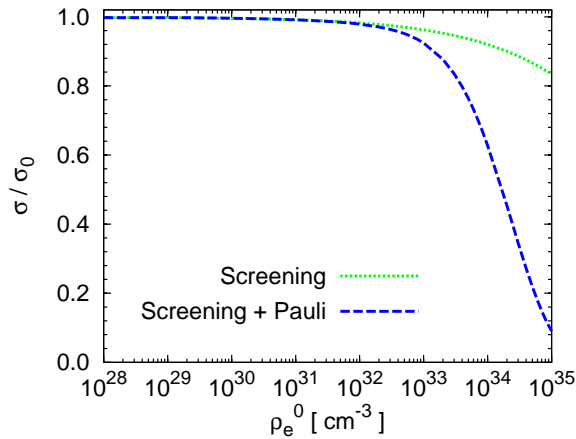}
\caption{
(Color online) 
The total cross sections for the 
$\nu_e + ^{56}$Fe $\rightarrow e^- + ^{56}$Co reaction 
as a function of the density of the environmental electrons. 
These are given as the ratio to the cross sections in the absence of the 
environmental electrons, $\sigma_0$. 
The meaning of each line is the same as in Fig. 1. } 
\label{fig:3}
\end{center}
\end{figure}
\begin{figure}
\begin{center}
\includegraphics[width=0.91\linewidth, clip]{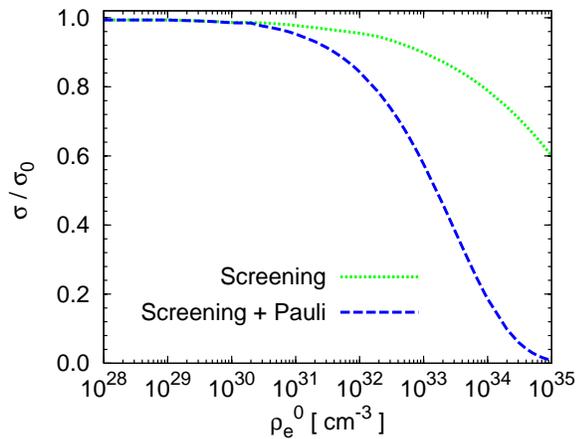}
\caption{
(Color online) 
The same as Fig. 3, but for the 
$\nu_e + ^{208}$Pb $\rightarrow e^- + ^{208}$Bi reactions. }
\label{fig:4}
\end{center}
\end{figure}

We find that 
the screening effect on the total cross sections 
is sensitive to the variation of the Coulomb potential due to the 
environmental electrons. 
For instance, 
the depth of the potential between the outgoing electron and 
$^{208}$Bi is 25.2 MeV in the absence of the 
environmental electrons. 
At $\rho^0_{e}=10^{31}, 10^{32}$ and 10$^{31}$ cm$^{-3}$, 
it is 25.08, 24.97, and 24.73 MeV, respectively. 
The corresponding change of the 
total cross section is 2.25, 4.48, and  10.1\%, 
respectively. 
Clearly, 
it is important to estimate carefully the Coulomb potential
in the high electron density region. 

\section{Summary}

The motion of the outgoing electron in the final state of a 
charged current neutrino-nucleus reaction 
is affected by surrounding environmental electrons in a supernova. 
We have discussed the electron screening as well as the Pauli blocking 
effects due to the environmental electrons on 
cross sections of the neutrino-nucleus reaction. 
For this purpose, we used the
Thomas-Fermi theory for the screening potential, 
and the DWBA with the Pauli blocking factor for the cross sections.
Our results for the 
$^{208}$Pb$(\nu_{e},e^{-})^{208}$Bi and 
$^{56}$Fe$(\nu_{e},e^{-})^{56}$Co reactions show that 
both the effects hinder the cross sections, especially at high
electron densities. 
We have also shown that the Pauli blocking effect is more significant 
in the former reaction than in the latter reaction due to the larger 
excitation energy for the Fermi transition.
The screening effect is also larger in the former reaction because of 
the larger atomic number. 

As far as we know, 
the screening and the Pauli 
blocking effects discussed in this paper have not been taken into 
account in the calculations for r-process nucleosynthesis. 
It would be of intriguing to see how these effects influence 
the r-process nucleosynthesis. 

\section*{Acknowledgment}
We thank Y. Kato for useful discussions on electron screening. 
We also acknowledge the 21st Century for Center of Excellence
(COE) Program ``Exploring New Science by Bridging Particle-Matter 
Hierarchy'' at Tohoku University. Discussions during the 4th COE 
symposium were useful to complete this work. 
This work was supported in part by the Grant-in-Aid for Scientific Research,
Contract No. 16740139 from the Japanese Ministry of Education,
Culture, Sports, Science, and Technology, in part 
by the U.S. National Science
Foundation Grant No. PHY-0555231, and in part by the University of
Wisconsin Research Committee with funds granted by the Wisconsin
Alumni Research Foundation.

\end{document}